\documentstyle[12pt]{article}
\title{Gaugino pair production at LHC  for the case of nonuniversal 
gaugino masses}
\author{\large S.I.~Bityukov~$^1$, N.V.~Krasnikov  \\[3mm]
\em Institute for Nuclear Research RAS, \\
\em Moscow, 117312, Russia  }
\date{}
\begin{document}
\maketitle
\begin{abstract}
We investigate $\tilde \chi^{\pm}_1 \tilde \chi^0_2$ pair production
at LHC (CMS) with subsequent decays into leptons for the case 
of nonuniversal gaugino masses. Visibility of signal by an excess
over SM background in $3l + no~jets + E^{miss}_T$ events
depends rather strongly on the relation between LSP mass 
$\tilde \chi^0_1$ and $\tilde \chi^{\pm}_1$ mass.
We also give some preliminary results on the investigation of squark
and gluino production at LHC for the case of nonuniversal gaugino
masses.
\end{abstract}
\vspace{1cm}
\bigskip

\noindent
\rule{3cm}{0.5pt}\\
$^1$~~Institute for High Energy Physics, Protvino, Russia

\section{Introduction}

One of the LHC goals is the discovery of the supersymmetry. 
In particular, it is very important to investigate a possibility
to discover nonstrongly interacting superparticles (sleptons,
higgsino, gaugino). In ref.\cite{1}
 (see, also references \cite{2,3}) the LHC gaugino discovery potential
has been investigated within the minimal SUGRA-MSSM framework where all 
sparticle masses are determined mainly by two parameters: $m_0$ (common
squark and slepton mass at GUT scale) and $m_{1 \over 2}$ (common
gaugino mass at GUT scale). 
The signature used for the search for gauginos at LHC is 
$3~isolated~leptons + no~jets + E^{miss}_T$ events. 
The conclusion of ref.~\cite{1}  is that LHC is able 
to detect gauginos with $m_{1 \over 2}$ up to 170~GeV and in some
cases (small $m_0$) up to 420~GeV.

In this paper we investigate the gaugino discovery potential of LHC for the
case of nonuniversal gaugino masses. Despite
the simplicity of the  SUGRA-MSSM framework it is a very particular
model. The mass formulae for sparticles in  SUGRA-MSSM model are derived
under the assumption that at GUT scale ($M_{GUT} \approx 2 \cdot 10^{16}$~GeV) 
soft supersymmetry breaking terms are universal. However, in general,
we can expect that real sparticle masses can differ in a drastic way 
from sparticle masses pattern of SUGRA-MSSM model due to many reasons,
see for instance refs.~\cite{4,5,6,7}.
Therefore, it is more appropriate to investigate the LHC SUSY discovery 
potential in a model-independent way \footnote{The early version of this 
study has been published in ref.~\cite{8}.}. The cross section 
for the $\tilde \chi^{\pm}_1 \tilde \chi^0_2$
chargino second neutralino pair production depends mainly on the mass of 
chargino which is approximately degenerate in mass with the second
neutralino $M(\tilde \chi^{\pm}_1) \approx M(\tilde \chi^0_2)$. 
The two lightest neutralino and the lightest chargino 
$(\tilde \chi^0_1, \tilde \chi^0_2, \tilde \chi^{\pm}_1)$ have, as largest
mixing components, the gauginos, and hence their masses are determined by
the common gaugino mass, $m_{1 \over 2}$. Within mSUGRA model
$M(\tilde \chi^0_1) \approx 0.4 m_{1 \over 2}$ and
$M(\tilde \chi^0_2) \approx M(\tilde \chi^{\pm}_1) 
\approx 2 M(\tilde \chi^0_1)$. 

The lightest chargino $\tilde \chi^{\pm}_1$ has several leptonic decay modes
giving an isolated lepton and missing energy:

three-body decay

\begin{itemize}

\item 
$\tilde \chi^{\pm}_1 \longrightarrow \tilde \chi^0_1  + l^{\pm} + \nu$,

\end{itemize}

two-body decays

\begin{itemize}

\item
$\tilde \chi^{\pm}_1  \longrightarrow  \tilde l^{\pm}_{L,R} + \nu$,

\hspace{16mm}  $\hookrightarrow \tilde \chi^0_1 + l^{\pm}$

\item
$\tilde \chi^{\pm}_1 \longrightarrow \tilde \nu_L + l^{\pm}$,

\hspace{16mm} $ \hookrightarrow \tilde \chi^0_1 + \nu$

\item
$\tilde \chi^{\pm}_1 \longrightarrow \tilde \chi^0_1 + W^{\pm}$.

\hspace{26mm} $ \hookrightarrow l^{\pm} + \nu$

\end{itemize}

Leptonic decays of $\tilde \chi^0_2$ give two isolated leptons and missing 
energy:

three-body decays

\begin{itemize}
\item 
$\tilde \chi^0_2 \longrightarrow \tilde \chi^0_1 + l^+ l^-$,

\item
$\tilde \chi^0_2 \longrightarrow \tilde \chi^{\pm}_1 + l^{\mp} + \nu$,

\hspace{16mm} $ \hookrightarrow \tilde \chi^0_1 + l^{\pm} + \nu$

\end{itemize}

two-body decay

\begin{itemize}
\item
$\tilde \chi^0_2 \longrightarrow \tilde l^{\pm}_{L,R} + l^{\mp}$.

\hspace{16mm} $ \hookrightarrow \tilde \chi^0_1 + l^{\pm}$

\end{itemize}

For relatively large $\tilde \chi^0_2$ mass there are two-body decays
$\tilde \chi^0_2 \longrightarrow \tilde \chi^0_1 h$,
$\tilde \chi^0_2 \longrightarrow \tilde \chi^0_1 Z$
which suppress three-body decay of $\tilde \chi^0_2$.
Direct production of $\tilde \chi^{\pm}_1 \tilde \chi^0_2$ followed
by leptonic decays of both gives three high $p_T$ isolated leptons
accompanied by missing energy due to escaping $\tilde \chi^0_1$'s and
$\nu$'s. These events do not contain jets except jets coming from
initial state radiation. Therefore the signature for 
$\tilde \chi^{\pm}_1 \tilde \chi^0_2$ pair production is 
$3l + no~jets +missing~energy$.

As mentioned above, this signature has been used in ref.~\cite{1}
for investigation of LHC gaugino discovery potential within mSUGRA model,
where gaugino masses $M(\tilde \chi^0_1)$, $M(\tilde \chi^0_2)$ are
determined mainly by a common gaugino mass $m_{1 \over 2}$ and 
$M(\tilde \chi^0_2) \approx 2.5 M(\tilde \chi^0_1)$. In our  
study we consider the general case when the relation between 
$M(\tilde \chi^{\pm}_1)$ and $M(\tilde \chi^0_1)$ is arbitrary.
We find that LHC gaugino discovery potential depends rather strongly on
the relation between $\tilde \chi^0_1$ and $\tilde \chi^0_2$ masses. 
For $M_{\tilde{\chi}_2^0} -M_{\tilde{\chi}_1^0} \geq M_Z $ the 
decay $\tilde{\chi}_2^0 \rightarrow 
\tilde{\chi}_1^0 Z$   dominates and due the real Z-boson in final state 
the signal 
is as a rule too small to be observable due to huge background. For 
$M_{\tilde{\chi}_2^0}$ closed 
to $M_{\tilde{\chi}_1^0}$ the leptons in final state are 
rather soft that also prevents the signal detection.  
We also give some preliminary results on the investigation of squark
and gluino production at LHC for the case of nonuniversal gaugino
masses.

\section{Simulation of detector response. Backgrounds}

Our simulations are made at the particle level with parametrized
detector responses based on a detailed detector simulation. 
To be concrete our estimates have been made for the CMS(Compact Muon 
Solenoid)  detector. The CMS detector simulation 
program CMSJET~3.2~\cite{9} is used.
The main aspects of the CMSJET relevant to our study are the 
following.

\begin{itemize}
\item
Charged particles are tracked in a 4 T magnetic field. 90 percent 
reconstruction  efficiency per charged track with $p_T > 1$~GeV within 
$|\eta| <2.5$ is assumed. 

\end{itemize}

\begin{itemize}
\item
The geometrical acceptances for $\mu$ and $e$ are $|\eta| <2.4$ and 2.5, 
respectively. The lepton number is smeared according to parametrizations 
obtained from full GEANT simulations. For a 10~GeV lepton the momentum 
resolution $\Delta p_T/p_T$ is better than one percent over the full $\eta$ 
coverage. For a 100~GeV lepton the resolution becomes $\sim (1 - 5) \cdot 
10^{-2}$ depending on $\eta$. We have assumed a 90 percent triggering 
plus reconstruction efficiency per lepton within the geometrical 
acceptance of the CMS detector.  

\end{itemize} 

\begin{itemize}
\item
The electromagnetic calorimeter of CMS extends up to $|\eta| = 2.61$. There 
is a pointing crack in the ECAL barrel/endcap transition region 
between $|\eta| = 1.478 - 1.566$ (6 ECAL crystals). The hadronic calorimeter 
covers $|\eta| <3$. The Very Forward calorimeter extends from $|\eta| <3$ 
to $|\eta| < 5$. Noise terms have been simulated with Gaussian distributions 
and zero suppression cuts have been applied. 
\end{itemize}

\begin{itemize}
\item
$e/\gamma$ and hadron shower development are taken into account by 
parametrization of the lateral and longitudinal profiles of showers. The 
starting point of a shower is fluctuated according to an exponential 
law.

\end{itemize}

\begin{itemize}
\item
For jet reconstruction we have used a slightly modified UA1 Jet Finding 
Algorithm, with a cone size of $\Delta R = 0.8$ and 25~GeV transverse 
energy threshold on jets. 

\end{itemize}
  
All SUSY processes with full particle spectrum, couplings,
production cross section and decays are generated with ISAJET~7.32,
ISASUSY~\cite{10}. The Standard Model backgrounds are generated 
with PYTHIA~5.7~\cite{11}. 

The following SM processes give the main contribution to the background:

\noindent
$WZ,~ZZ,~t \bar t,~Wtb,~Zb \bar b,~b \bar b$. 
In this paper we use the results of
the background simulation of ref.~\cite{1}. Namely following ref.~\cite{1}
we require 3 isolated leptons with $p^l_T > 15~GeV$ in
$|\eta^l| < 2.4~(2.5)$ for muons (electrons)
and with the same-flavour opposite-sign leptons.
As an lepton isolation criterium we require the absence of charged tracks
with $p_T > 1.5~GeV$ in a cone $R = 0.3$ around lepton. We require
also the absence of jets with $E^{jet}_T > 30~GeV$ in $|\eta^l| < 3$.
The last requirement is that the two same-flavour opposite-sign lepton
invariant mass $M_{l^+l^-} < 81~GeV$. Lepton isolation is useful  for the  
suppression of the  background events with leptons originating 
from semileptonic decays of b-quarks. The central jet veto requirement 
allows to get rid of the internal SUSY background coming from 
$\tilde{g}$ and $\tilde{q}$ cascade decays, which otherwise overwhelms 
$\tilde{\chi}^{\pm}_1$ $\tilde{\chi}^0_2$ direct production. Also this 
cut reduces $t\bar{t}$, $Wtb$, $Zb\bar{b}$, $b\bar{b}$ SM background.
 
For such set of cuts the background cross section
$\sigma_{back} = 10^{-2}~pb$~\cite{1} that corresponds to the number
of background events $N_b = 10~(100)$ for total  luminosity
$L = 10^3~(10^4)~pb^{-1}$. See for details ref.\cite{1}. If we refuse from 
the $M_Z$-cut, namely if we refuse from the requrement that the two 
same-flavour opposite-sign lepton invariant mass $M_{l^+l^-} <81$~GeV, then 
the background cross section is $\sigma_{back} = 0.11~pb$~\cite{1}.    

\section{Results}

The results of our calculations are presented in Tables 1-9. In
estimation of the LHC(CMS) gaugino discovery potential we have used
the significance determined as $S_{12} = \sqrt{N_s + N_b} - \sqrt{N_b}$
which is appropriate for the estimation of discovery potential in the case
of future experiments~\cite{12}. For the comparison we also give the 
values of often used significance determined as 
$S = \frac{N_S}{\sqrt{(N_S+N_B)}}$. 
Here $N_s = \sigma_s \cdot L$ is the number
of signal events and $N_b = \sigma_b \cdot L$ is the number of background 
events for a given total luminosity $L$. As it follows from our results 
for given value of chargino mass $M(\tilde \chi^{\pm}_1)$ the number of
signal events depends rather strongly on the mass of the lightest 
superparticle $M(\tilde \chi^0_1)$ and for 
$M(\tilde \chi^0_1) \ge 0.7 M(\tilde \chi^{\pm}_1)$ signal is too
small to be observable. For small LSP masses $M(\tilde{\chi}_1^0)$ two body 
decay $\tilde{\chi}^0_2 \rightarrow \tilde{\chi}^0_1~Z$ dominates and due to 
$M_Z$-cut signal is as a rule too small to be observable. However if we 
refuse from $M_Z$ cut in some cases it is possible to detect 
two body decay mode $\tilde{\chi}_2^0 \rightarrow \tilde{\chi}_1^0(Z 
\rightarrow l^+l^-)$. As an illustration consider several examples 
wich correspond to total luminosity $L = 3 \cdot 10^{4}$ $pb^{-1}$ 
(for such luminosity the number of baskground events is expected to be 
equal 3300).

A. $M_{\tilde{q}} =M_{\tilde{l}} = M_{\tilde{g}}  = 2$ TeV. $\tan(\beta) =5$,
$M(\tilde{\chi}^0_2) \approx M(\tilde{\chi}^{\pm}_1) = 104$~GeV, 
$M(\tilde{\chi}^0_1) = 11~GeV, N_{ev} = 1921, S_{12} =14, S = 26$.

B. $M_{\tilde{q}} =M_{\tilde{l}} = M_{\tilde{g}}  = 2$ TeV. $\tan(\beta) =5$,
$M(\tilde{\chi}^0_2) \approx M(\tilde{\chi}^{\pm}_1) = 126$~GeV, 
$M(\tilde{\chi}^0_1) = 21~GeV, N_{ev} = 924, S_{12} =8, S = 14$.

C. $M_{\tilde{q}} =M_{\tilde{l}} = M_{\tilde{g}}  = 500$~GeV. $\tan(\beta) =5$,
$M(\tilde{\chi}^0_2) \approx M(\tilde{\chi}^{\pm}_1) = 122$~GeV, 
$M(\tilde{\chi}^0_1) = 26~GeV, N_{ev} = 744, S_{12} =6, S = 11$.

D. $M_{\tilde{q}} =M_{\tilde{l}} = M_{\tilde{g}}  = 1$ TeV. $\tan(\beta) =5$,
$M(\tilde{\chi}^0_2) \approx M(\tilde{\chi}^{\pm}_1) = 124$~GeV, 
$M(\tilde{\chi}^0_1) = 32~GeV, N_{ev} = 864, S_{12} =7.1, S = 13$.

\section{Squark and gluino production for the case of nonuniversal
gaugino masses}

Here we give first preliminary results on the squark and gluino
production for the case of nonuniversal gaugino masses~\cite{13,14}.
We investigated 3 cases:

\begin{itemize}
\item[A.] $m_{\tilde q} \ll m_{\tilde g}$,

\item[B.] $m_{\tilde q} \gg m_{\tilde g}$,

\item[C.] $m_{\tilde q} \ge m_{\tilde g}$, $m_{\tilde q} \sim m_{\tilde g}$.

\end{itemize}

We used the signature $n \ge 2$ {\it jets} + $E^{miss}_T$ for the
supersymmetry search. As for the signature
$n \ge 2$ {\it jets} + $n \ge 1$ {\it leptons} + $E^{miss}_T$, it
depends on the relation among 
$\tilde \chi^0_2, \tilde \chi^0_1, \tilde q$ and $\tilde g$ masses.
For the instance, for 
$m_{\tilde \chi^0_2} \ge min(m_{\tilde g}, m_{\tilde q})$ 
there are no cascade decays of
$\tilde \chi^0_2$ and, hence, the signature with $n \ge 1$ {\it leptons}
is not essential.

We have found that for the case of arbitrary relations among gaugino masses
the number of signal events for the signature
$n \ge 2$ {\it jets} + $E^{miss}_T$  depends rather strongly on the relation
among $m_{\tilde g}$, $m_{\tilde q}$ and $m_{\tilde \chi^0_1}$.
For $m_{\tilde \chi^0_1}$ closed to $min(m_{\tilde g}, m_{\tilde q})$
the perspectives of SUSY detection become very problematic.
For instance, for $min(m_{\tilde g}, m_{\tilde q}) \ge 1~TeV$ and 
$m_{\tilde \chi^0_1} \ge 0.75 \cdot min(m_{\tilde g}, m_{\tilde q})$
it is extremely difficult or even impossible to detect SUSY using the channel 
$n \ge 2$ {\it jets} + $E^{miss}_T$.

As a concrete example consider the SUSY detection for 
$m_{\tilde q} = 1550~GeV$, $m_{\tilde g} = 1500~GeV$. For cut with
$n_{jet} \ge 3$, $p_{T_{jet1}} \ge 350~GeV$, $p_{T_{jet2}} \ge 290~GeV$, 
$p_{T_{jet3}} \ge 230~GeV$, $E^{miss}_T \ge 1200~GeV$ for
luminosity $L = 10^5pb^{-1}$ the number of background events is 
35 whereas the number of signal events is 334,
66, 9, 5 for $m_{\tilde \chi^0_1} = 250~GeV,~750~GeV,~1125~GeV$ and 
$1350~GeV$, correspondingly.

\section{Conclusion}

In this paper we have presented the results of the calculations for 
$\tilde \chi^{\pm}_1 \tilde \chi^0_2$ pair production at LHC (CMS) with their 
subsequent decays into leptons for the case of nonuniversal gaugino masses.
We have found that the visibility of signal by an excess over SM background
in $3l + no~jets + E^{miss}_T$ events depends rather strongly on the relation 
between LSP mass $\tilde \chi^0_1$ and chargino $\tilde \chi^{\pm}_1$ mass. 
For relatively heavy LSP mass  
$M_{\tilde{\chi}^0_1} \geq M_{\tilde{\chi}^{\pm}_1}$ signal is too small 
to be observable.   Also for small values of LSP mass 
$M_{\tilde{\chi}^0_1}$ two body decay $\tilde{\chi}^0_2  \rightarrow 
\tilde{\chi}^0_1~Z $ 
complicates the observation of the signal.
For total luminosity $L = 3 \cdot 10^4pb^{-1}$ signal could be 
observable for chargino mass $M(\tilde \chi^{\pm}_1)$ up to 
150~GeV .

\begin{center}
 {\large \bf Acknowledgments}
\end{center}

\par
We are  indebted to I.N.~Semeniouk for his help in writing the code of 
the events selections. This work has been supported by RFFI grant 99-02-16956.


\newpage

\begin{table}[h]
\small
     \caption{The number of events $N_{ev}$ and 
significances $S_{12}$, $S$ for  
$M(\tilde \chi^0_2)~=~104~GeV,~ M(\tilde q) = 2~TeV$,
$L~=~3 \cdot 10^4~pb^{-1}$
and for different LSP masses $M(\tilde \chi^0_1)$.}
    \label{tab.1}
\begin{center}
\begin{tabular}{|l|l|l|l|l|l|l|l|l|l|l|l|l|l| }
\hline
$ M(\tilde{\chi}^0_1)~(GeV)$ & 11 & 14 & 21 & 26 & 31 & 36 & 41 & 46  & 51
 & 56 & 61 & 71 & 81 \\
\hline
$N_{ev}$ & 181 & 662 & 571 & 330 & 331 & 241 & 240 & 182 & 179 & 181 &121 
&21 & 4 \\
\hline
$ S_{12} $  & 4.6 & 13.8 & 12 & 7.7 & 7.7 & 6.0 & 6.0 & 4.6 & 4.6 & 4.6 & 
3.2 & 0.6 & 0.12 \\
\hline
$ S $ & 8.2 & 21.4 & 19.2 & 13.2 & 13.2 &  10.4 & 10.4 & 8.3 & 8.3  & 
8.3 & 6.2 & 1.2 & 0.2 \\ 
\hline
\end{tabular}
\end{center}
\end{table}

\begin{table}[h]
\small
    \caption{The number of events $N_{ev}$ and 
significances $S_{12}$ , $S$ for  
$M(\tilde \chi^0_2)~=~126~GeV$, $M(\tilde q) = 2~TeV$,
$L~=~3 \cdot 10^{4}~pb^{-1}$
and for different LSP masses $M(\tilde \chi^0_1)$.}
    \label{tab:Tab.2}
\begin{center}
\begin{tabular}{|l|l|l|l|l|l|l|l|l| }
\hline
$ M(\tilde \chi^0_1)~(GeV)$ & 11 & 21 & 34 & 38 & 46  & 66 & 76 & 86  \\
\hline
$N_{ev}$  & 31 & 28 & 73 & 321 & 211  & 161 &  136 & 79 \\
\hline
$ S_{12} $ & 0.85 & 0.76 & 1.9 & 7.6  & 3.1 & 4.2 &  3.5 & 2.1 \\
\hline
$S$ & 1.6 & 1.5 & 3.8 & 12.8 & 9.3 & 7.6 & 6.4 & 4.0 \\
\hline
\end{tabular}
\end{center}
\end{table}

\begin{table}[h]
\small
    \caption{The number of events $N_{ev}$ and 
significances $S_{12}$, $S$ for 
$M(\tilde \chi^0_2)~=~150~GeV$, $M(\tilde q) = 2~TeV$,
$L~=~3 \cdot 10^{4}~pb^{-1}$
and for different LSP masses $M(\tilde \chi^0_1)$.}
    \label{tab:Tab.3}
\begin{center}
\begin{tabular}{|l|l|l|l|l|l|l|l|l| }
\hline
$ M(\tilde \chi^0_1)~(GeV)$ & 11 & 41 & 56 & 61 & 66 & 71 & 81 & 91 \\
\hline
$N_{ev}$             & 15 & 10 & 35 & 161 & 102 & 94 & 65 & 42  \\
\hline
$ S_{12} $       & 0.4 & 0.3 & 1 & 4.2 & 2.8 & 2.5 & 1.8 & 1.2  \\
\hline
$S$ &    0.9 & 0.6 & 1.9 & 7.5 & 5.1 & 4.7 & 3.4 & 2.3 \\
\hline
\end{tabular}
\end{center}
  \end{table}

\begin{table}[h]
\small
    \caption{The number of events $N_{ev}$ and 
significances  $S_{12}$, $S$  for  
$M(\tilde \chi^0_2)~=~101~GeV$, $M(\tilde q) = 1000~GeV$,
$L~=~3 \cdot 10^{4}~pb^{-1}$
and for different LSP masses $M(\tilde \chi^0_1)$.}
    \label{tab:Tab.4}
\begin{center}
\begin{tabular}{|l|l|l|l|l|l|l|l|l| }
\hline
$ M(\tilde \chi^0_1)~(GeV)$ & 10 & 20 & 25 & 30 & 35 & 45 & 50 & 60 \\
\hline
$N_{ev}$     & 451   & 571 & 439 & 239 & 302 & 298 & 178 & 61  \\
\hline
$ S_{12} $     & 9.9 & 12.1 & 9.9 & 5.5 & 7.2 & 7.2 & 4.4 & 1.1 \\
\hline
$S$  & 16.5 & 19.2 & 16.5 & 10.4 & 12.1 & 12.1 & 8.2 & 3.3  \\
\hline
\end{tabular}
\end{center}
  \end{table}

\begin{table}[h]
\small
     \caption{The number of events $N_{ev}$ and 
significances $S_{12}$ , $S$ for  
$M(\tilde \chi^0_2)~=~124~GeV,~ M(\tilde q) = 1~TeV$,
$L~=~3 \cdot 10^4~pb^{-1}$
and for different LSP masses $M(\tilde \chi^0_1)$.}
    \label{tab.5}
\begin{center}
\begin{tabular}{|l|l|l|l|l|l|l|l|l|l| }
\hline
$ M(\tilde \chi^0_1)~(GeV)$ & 10 & 31 & 34 & 36 & 40 & 50 & 60 & 70 & 80 \\
\hline
$N_{ev}$       & 31 & 24 & 353 & 305 & 207 & 166 & 134 & 92 & 63 \\
\hline
$ S_{12} $    & 0.9 & 0.7 & 8.3 & 7.3 & 5.2 & 4.2 & 3.4 & 2.4 & 1.7 \\
\hline
$S$          & 1.7 & 1.4 & 13.8 & 12.4 &  9.2 & 7.6 & 6.4 & 4.7 & 3.3 \\
\hline
\end{tabular}
\end{center}
\end{table}

\begin{table}[h]
\small
     \caption{The number of events $N_{ev}$ and 
significances  $S_{12}$ , $S$ for  
$M(\tilde \chi^0_2)~=~151~GeV,~ M(\tilde q) = 1~TeV$,
$L~=~3 \cdot 10^4~pb^{-1}$
and for different LSP masses $M(\tilde \chi^0_1)$.}
    \label{tab.6}
\begin{center}
\begin{tabular}{|l|l|l|l|l|l|l|l|l|l| }
\hline
$ M(\tilde \chi^0_1)~(GeV)$ & 10 & 60 & 62 & 65 & 70 & 80 & 90 & 100 & 110 \\
\hline
$N_{ev}$     & 3 & 19 & 157 & 123 & 98 & 93 & 85 & 83 & 17 \\
\hline
$ S_{12} $    & 0.005 & 0.03 & 4.0 & 3.3 & 2.6 & 2.4 & 2.2 & 1.9 & 0.5 \\
\hline
$S$    & 0.01 & 0.06 & 7.3 & 6.0 & 4.8 & 4.7 & 4.3 & 3.8 & 1.0 \\  
\hline
\end{tabular}
\end{center}
\end{table}

\begin{table}[h]
\small
     \caption{The number of events $N_{ev}$ and 
significances $S_{12}$, $S$ for  
$M(\tilde \chi^0_2)~=~98~GeV,~ M(\tilde q) = 500~GeV$,
$L~=~3 \cdot 10^4~pb^{-1}$
and for different LSP masses $M(\tilde \chi^0_1)$.}
    \label{tab.7}
\begin{center}
\begin{tabular}{|l|l|l|l|l|l|l|l| }
\hline
$ M(\tilde \chi^0_1)~(GeV)$ & 4 & 10 & 19 & 29 & 39 & 49 & 59  \\
\hline
$N_{ev}$      & 151 & 619 & 211 & 180 & 178 & 148 & 60  \\
\hline
$ S $           & 4.4 & 13.2 & 5.3 & 4.6 & 4.6 & 3.9 & 1.7  \\
\hline
$S$ & 7.2 & 20.9 & 9.4 & 8.3 & 8.3 & 7.2 & 3.3 \\
\hline
\end{tabular}
\end{center}
\end{table}

\begin{table}[h]
\small
     \caption{The number of events $N_{ev}$ and 
significances $S_{12}$ , $S$ for  
$M(\tilde \chi^0_2)~=~122~GeV,~ M(\tilde q) = 500~GeV$,
$L~=~3 \cdot 10^4~pb^{-1}$
and for different LSP masses $M(\tilde \chi^0_1)$.}
    \label{tab.8}
\begin{center}
\begin{tabular}{|l|l|l|l|l|l|l|l|l| }
\hline
$ M(\tilde \chi^0_1)~(GeV)$ & 9 & 29 & 35 & 39 & 43 & 51 & 58 & 75 \\
\hline
$N_{ev}$      & 15 & 26 & 191 & 155 & 129 & 124 & 71 & 63 \\
\hline
$ S_{12} $      & 0.4 & 0.7 & 4.8 & 3.5 & 3.5 & 3.3 & 1.9 & 1.7 \\
\hline
$S$      & 0.9 & 1.4 & 9.0 & 7.3 & 6.2 & 6.0 & 3.6 & 3.3 \\ 
\hline
\end{tabular}

\end{center}
\end{table}
        
\begin{table}[h]
\small
     \caption{The number of events $N_{ev}$ and 
significances $S_{12}$ , $S$ for  
$M(\tilde \chi^0_2)~=~146~GeV,~ M(\tilde q) = 500~GeV$,
$L~=~3 \cdot 10^4~pb^{-1}$
and for different LSP masses $M(\tilde \chi^0_1)$.}
    \label{tab.9}
\begin{center}
\begin{tabular}{|l|l|l|l|l|l|l| }
\hline
$ M(\tilde \chi^0_1)~(GeV)$ & 9 & 53 & 56 & 59 & 65 & 69  \\
\hline
$N_{ev}$                 & 18 & 6 & 109 & 105 & 52 & 48  \\
\hline
$ S_{12} $         & 0.3 & 0.1 & 2.9 & 2.8 & 1.4 & 1.4  \\
\hline
$S$    & 0.6 & 0.2 & 5.4 & 5.3 & 2.8 & 2.6 \\
\hline
\end{tabular}
\end{center}
\end{table}


\begin{thebibliography}{99}

\bibitem{1} I.Iashvili, A.Kharchilava and K.Mazumdar, {\it Study
of $\tilde \chi^+_1 \tilde \chi^0_2$ Pair Production with the CMS Detector
at LHC}, CMS NOTE 1997/007.
\bibitem{2} R.Barbieri et al., Nucl.Phys. {\bf B367}(1993)28.
\bibitem{3} H.Baer, C.Chen, F.Paige and X.Tata,
Phys.Rev. {\bf D50}(1994)2148.
\bibitem{4} V.S.Kaplunovsky and J.Louis, Phys.Lett. {\bf B306}(1993)269.
\bibitem{5} N.Polonsky and A.Pomarol, Phys.Rev.Lett. {\bf 73}(1994)2292.
\bibitem{6} N.V.Krasnikov and V.V.Popov,
{\it PLANCSUSY~-~new program for SUSY masses calculations:
from Planck scale to our reality,} Preprint INR 976TH/96.
\bibitem{7} C.Kolda and J.March-Russel, Phys.Rev.{\bf D55}(1997)4252.
\bibitem{8} S.I.Bityukov and N.V.Krasnikov, Gaugino production at LHC(CMS), 
hep-ph/9810294, \\
to be published in proceedings of the international 
conference on high energy physics, 
Dubna, 9-13 July 1998.
\bibitem{9} S.Abdullin, A.Khanov and N.Stepanov, {\it CMSJET 3.2,
CMSJET 3.5,} CMS Note CMS TN/94-180.
\bibitem{10}  H.Baer, F.Paige, S.Protopesku and X.Tata, {\it Simulating 
Supersymmetry with ISAJET 7.0/ISASUSY 1.0,} Florida State University Preprint 
EP-930329(1993).
\bibitem{11}T.Sjostrand, {PYTHIA 5.7 and ISAJET 7.4, Physics and Manual,}
CERN-TH.7112/93.
\bibitem{12} S.I.Bityukov and N.V.Krasnikov, {\it Towards the Observation
of Signal over Background in Future Experiments}, INR Preprint 0945a/98,
also physics/9808016. 
\bibitem{13} N.V.Krasnikov, {\it It is always possible to discover
supersymmetry broken at TeV scale at LHC ?}, hep-ph/9001398. 
\bibitem{14} S.I.Bityukov and N.V.Krasnikov, to be published. 
\end{thebibliography}
\end{document}